\documentclass[useAMS,usenatbib]{mn2e}
\usepackage{graphicx,amssymb,color,wasysym}
\usepackage[normalem]{ulem}
\newcommand{\rev}{ }

\title[Smallest drivers of WD pollution]
{The smallest planetary drivers of white dwarf pollution}

\author[]{Dimitri Veras$^{1,2,3}$\thanks{E-mail: dimitri.veras@colorado.edu},
Aaron J. Rosengren$^{4,5}$ 
\\
$^{1}$Centre for Exoplanets and Habitability, University of Warwick, Coventry CV4 7AL, UK
\\
$^{2}$Centre for Space Domain Awareness, University of Warwick, Coventry CV4 7AL, UK
\\
$^{3}$Department of Physics, University of Warwick, Coventry CV4 7AL, UK
\\
$^{4}$Center for Astrophysics and Space Sciences, Department of Physics, UC San Diego, 92093, La Jolla, CA, USA
\\
$^{5}$Department of Mechanical and Aerospace Engineering, UC San Diego, 92093, La Jolla, CA, USA
}

\pubyear{2022}

\begin{document}
\label{firstpage}
\pagerange{\pageref{firstpage}--\pageref{lastpage}}
\maketitle

\begin{abstract}
Many potential mechanisms for delivering planetary debris to within a few Roche radii of white dwarfs rely on gravitational scattering events that feature perturbers which are giant planets or terrestrial planets. However, the population of these planets orbiting white dwarfs is still unknown, and for a substantial fraction of white dwarfs the largest planetary survivors of stellar evolution may be sub-terrestrial mass minor planets. Here, we attempt to identify the smallest mass perturbers that could pollute white dwarfs. Through computationally expensive numerical simulations of both unstable and stable configurations of minor planets, we find that this critical lower bound equals approximately one Luna mass ($1M_{\leftmoon} \approx 10^{-1}M_{\mars} \approx 10^{-2}M_{\oplus} \approx 10^2M_{\rm Ceres}$). Further, we find that as this mass limit is approached from above, the typical cooling age at which white dwarf pollution occurs increases. Consequently, there is a two order-of-magnitude range of perturber masses between Earth and its moon that has remained largely unexplored in white dwarf pollution studies, despite the potential formation of thousands of such Luna-sized objects in these systems. 
\end{abstract}


\section{Introduction}

Polluted white dwarfs -- in which planetary debris has ``polluted" the otherwise pristine H or He stellar atmospheres -- are ubiquitous \citep{zucetal2010,koeetal2014,couetal2019} and exhibit a startling variety of chemistry, abundances, and ages. In a given system, pollution can be explained chemically by the ingestion of asteroids {\rev \citep{zucetal2007,ganetal2012,doyetal2019,bonetal2022,bucetal2022,curetal2022}}, comets \citep{faretal2013,radetal2015,genetal2017,xuetal2017,hosetal2020}, or moons \citep{doyetal2021,kleetal2021}. The pollution can occur around both very young white dwarfs, with cooling ages under hundreds of Myr \citep{faretal2016,izqetal2021}, or very old white dwarfs, with cooling ages approaching 10 Gyr \citep{holetal2018,bloxu2022,elmetal2022}.

The diversity of this population of polluted white dwarfs casts doubt on a one-size-fits-all model for the delivery of planetary debris to the white dwarf atmosphere. The debris itself, having survived the star's giant branch phases, must have originated from distances of at least a few au \citep{musvil2012,madetal2016,ronetal2020}. The debris could have once been part of a fully formed minor or major planet, or smaller object such as a boulder or cobble, that was broken up either during the giant branch \citep{veretal2014b,versch2020} or white dwarf \citep{jura2003,broetal2017,mcdver2021} phases. Primordial dust or sand, however, would not have survived the giant branch phases of stellar evolution \citep{bonwya2010,donetal2010,zotver2020}.

From a distance of a few au, the polluting material must traverse a path to the white dwarf -- within about a couple solar radii from the star's centre -- where a disc is formed that is eventually accreted by the photosphere \citep{cunetal2022}. Traversing this path could be achieved solely due to radiative drag from the white dwarf \citep{veretal2022} although much more commonly proposed scenarios involve at least one massive perturbing object (see Fig. 6 of \citealt*{veras2021} for a list). In particular, the process of a planet perturbing an asteroid onto a highly eccentric orbit, followed by the breakup of the asteroid at its orbital pericentre, and subsequent contraction, has been investigated with increasing layers of detail \citep{debetal2012,veretal2014a,veretal2015b,malper2020,ocolai2020,lietal2021,zhaetal2021,broetal2022}.

The planets that were invoked in these studies were not based on actual planets known to orbit white dwarfs, which so far number only five or six \citep{thoetal1993,sigetal2003,luhetal2011,ganetal2019,vanetal2020,blaetal2021,gaietal2022}. Further, well over half of the known exoplanets will not survive to the white dwarf phase \citep{maletal2020a,maletal2020b,maletal2021}; the planets that will survive are, nevertheless, often beyond our current observational capabilities to detect. This observational bias is particularly acute for small, terrestrial planets, which outnumber large, giant planets by several multiples \citep{casetal2012,heretal2019}.

Hence, in pollution delivery scenarios involving gravitational scattering, the properties of the perturbers remain very poorly constrained. In principle, in contrast to common modelling choices, the perturbers do not need to be as massive as terrestrial planets. In fact, sub-terrestrial mass bodies may be common. In our own Solar system, thousands of these objects could have once existed; \cite{nesvok2016} claimed that $1000-4000$ Plutos fit origin scenarios well. 

Furthermore, recent investigations are revealing that smaller perturbers can more easily pollute white dwarfs at late ages \citep{frehan2014,musetal2018,veretal2021,ocoetal2022}. However, as the perturber masses approach zero, {\rev they eventually become too small to dynamically affect pollution}. Hence, there should be a limiting lower perturber mass below which pollution delivery is effectively unrealistic.

In this paper, we investigate sub-terrestrial mass scattering of debris into white dwarfs with the specific purpose of identifying the least-massive possible perturber. We explore this question with numerical simulations (Section 2). These are computationally expensive because of the inclusion of both the number of massive objects that perturb all others in the system, and the $\sim$10 Gyr-long timescales, not to mention the expansive potential parameter space to explore. Nevertheless, we obtain a well-constrained result to within a factor of a few in mass, as discussed in Section 3. We then summarise in Section 4.

\section{Numerical simulations}

\subsection{Initial conditions}

In order to find the least massive perturber that can pollute a white dwarf, we sought to construct the most dynamically active simulations that feature these perturbers.  Although the perturbers do not necessarily have to become unstable themselves \citep{musetal2018,ocoetal2022}, doing so may certainly facilitate the pollution process due to the radial incursions that are often generated by these instabilities \citep{musetal2014,vergan2015,veretal2016,veras2016b}.

\subsubsection{Perturber properties}

Consequently, we constructed sets of slightly-to-moderately packed perturbers that were likely to undergo instability during the white dwarf phase as a result of stellar mass loss during the giant branch phases \citep{debsig2002,veretal2013,veras2016a}. We did not intend to impose maximally ``packed" configurations \citep{smilis2009,funetal2010,fanmar2013,puwu2015,weietal2022} because we also hoped to explore perturber configurations that changed with time but did not necessarily experience instability.

We then combined these considerations with the maximum number of perturbers that could be included in a single simulation and remain computationally feasible. This number is 10, although we also performed a select few simulations with 15, 20, and 30 perturbers. In addition to the perturbers, in each system we included 15 test particles; these represent the potentially polluting debris. The test particles did not gravitationally affect one another or the perturbers, whereas each perturber affected every other object in the system.

To maximise the prospects for pollution, we placed the innermost perturber at 3 au along the main sequence {\rev for most of our simulations}. This value represents an approximate critical distance within which perturbers would be engulfed during the giant branch phases \citep{musvil2012,madetal2016,ronetal2020}. {\rev However, we recognize that this value is model-dependent and uncertain, and may be smaller. Hence, we also perform a set of simulations with the innermost perturber at 2~au along the main-sequence.} 

After having established the number of perturbers and their innermost extent, we {\rev then set} their masses and separations. {\rev We explored perturber masses of $M_{\rm per} = \left[0.5, 1, 2, 3, 4, 5, 7, 10\right] M_{\leftmoon}$}, where $1M_{\leftmoon}$ represents the mass of Earth's moon, which we will henceforth refer to as Luna. {\rev Also, for our fiducial simulations, we} set an initial semi-major axis range of $3-6$ au, with perturber values equally spaced within this range. Because our primary goal was to identify the critical lower limit of $M_{\rm per}$, in each simulation each perturber was given the same mass. Adopting a mass distribution would not only have been computationally unfeasible, but also not too revealing because differential perturber masses would generate less pollution than a configuration where the larger perturber in the distribution was duplicated 10 times.

Regarding the other orbital parameters of the perturbers, in packed configurations, changing the eccentricity of the perturbers often serves only to alter the instability timescale \citep{gralis2021}. In this respect, we did not expect this eccentricity choice to significantly affect our results. We chose eccentricities randomly from a uniform distribution in the interval $0.00-0.05$, and inclinations randomly from a uniform distribution in the interval $0^{\circ}-5^{\circ}$. The inclinations are measured from a fixed plane that intersects the star, which can be imagined to coincide with its equator. All initial orbital angles of the perturbers (argument of pericentre, longitude of ascending node, mean anomaly) were randomly sampled from uniform distributions across their entire allowable ranges.

\subsubsection{Test particle properties}

The next question we had to address was where to place the test particles. Similar to the perturbers, debris within a few au of the host star should not survive the giant branch phases\footnote{We did not model here the effects of radiation or common envelope evolution. Immediately after the star has contracted into a white dwarf, the white dwarf's luminosity could drag material inwards to the star itself \citep{veretal2022}. The only objects that could survive common envelope evolution, within the envelope itself, are much larger than Jupiter \citep{norspi2013,chaetal2021,yaretal2022}.}. Based on this premise, we also set the innermost semi-major axis of the debris to 3 au {\rev for most simulations}. Then, we embedded our test particles within our sea of perturbers, at a range of $3-6$ au. However, in contrast to the perturbers, the initial semi-major axes of the text particles were randomly selected. In some simulations we instead used the {\rev ranges $2-4$ au and $6-9$ au}, to help quantify the difference in outcomes when most of the debris is {\rev present at different locations}. We included 15 test particles per simulation because we found that this value provided the most efficient use of our computational resources.

The other initial orbital parameters of the particles were chosen to roughly mimic parameter ranges for exo-asteroid belts, which represent one of the most commonly assumed classes of polluters. In this respect, the initial eccentricities and inclinations of the test particles were randomly sampled in uniform ranges of $0.0-0.3$ and $0^{\circ} - 40^{\circ}$ respectively. Their orbital angles were randomly chosen in a similar way to those of the perturbers. 

\subsubsection{Other simulation properties}

We evolved the systems for 11 Gyr from the start of the giant branch phases by using the RADAU $N$-body integrator within a heavily modified version of the {\tt Mercury}-based software package \citep{chaetal1999}, where the modification allows for the simultaneous evolution of planetary systems and their host stars. This modification was first implemented in \cite{veretal2013} but then improved upon in \cite{musetal2018}. We adopted a RADAU integrator tolerance of $10^{-11}$, and, for most simulations, a main-sequence host star mass of $2M_{\odot}$, which is reflective of the typical progenitor mass of the current white dwarf population \citep{treetal2016,cumetal2018,elbetal2018,mccetal2020,barcha2021}. 

These choices led to demanding computations, prompting us to select a parameter output frequency of 1 Myr, although instability detection with respect to collisions between bodies or escape from the system was flagged at the timestep at which such events occurred. The white dwarf's physical radius was artificially inflated to $1R_{\odot}$ to mimic its Roche sphere radius, although in reality this value can vary typically by a factor of 2 depending on the properties of the object being accreted \citep{zhaetal2021,veretal2022}; direct impact onto the white dwarf photosphere is rare from an orbital perspective \citep{veretal2021} and requires high internal strength and resistance to sublimation \citep{mcdver2021,steetal2021}.

Our running time of 11 Gyr allowed us to explore very old polluted white dwarfs while also including the giant branch phases of evolution. These giant branch phases are important dynamically because of the changes they impose on the structure and stability of planetary systems \citep{veras2016a}. The stellar evolution prescriptions we adopted were from \cite{huretal2000}. According to this prescription, a $2.0M_{\odot}$ main-sequence star becomes a white dwarf after about 1.5 Gyr, retaining only about 32 per cent of its original mass.

\begin{figure*}
\centerline{\Huge \underline{A fiducial example with $4M_{\leftmoon}$ perturbers}}
\centerline{}
\centerline{
\includegraphics[width=8.5cm]{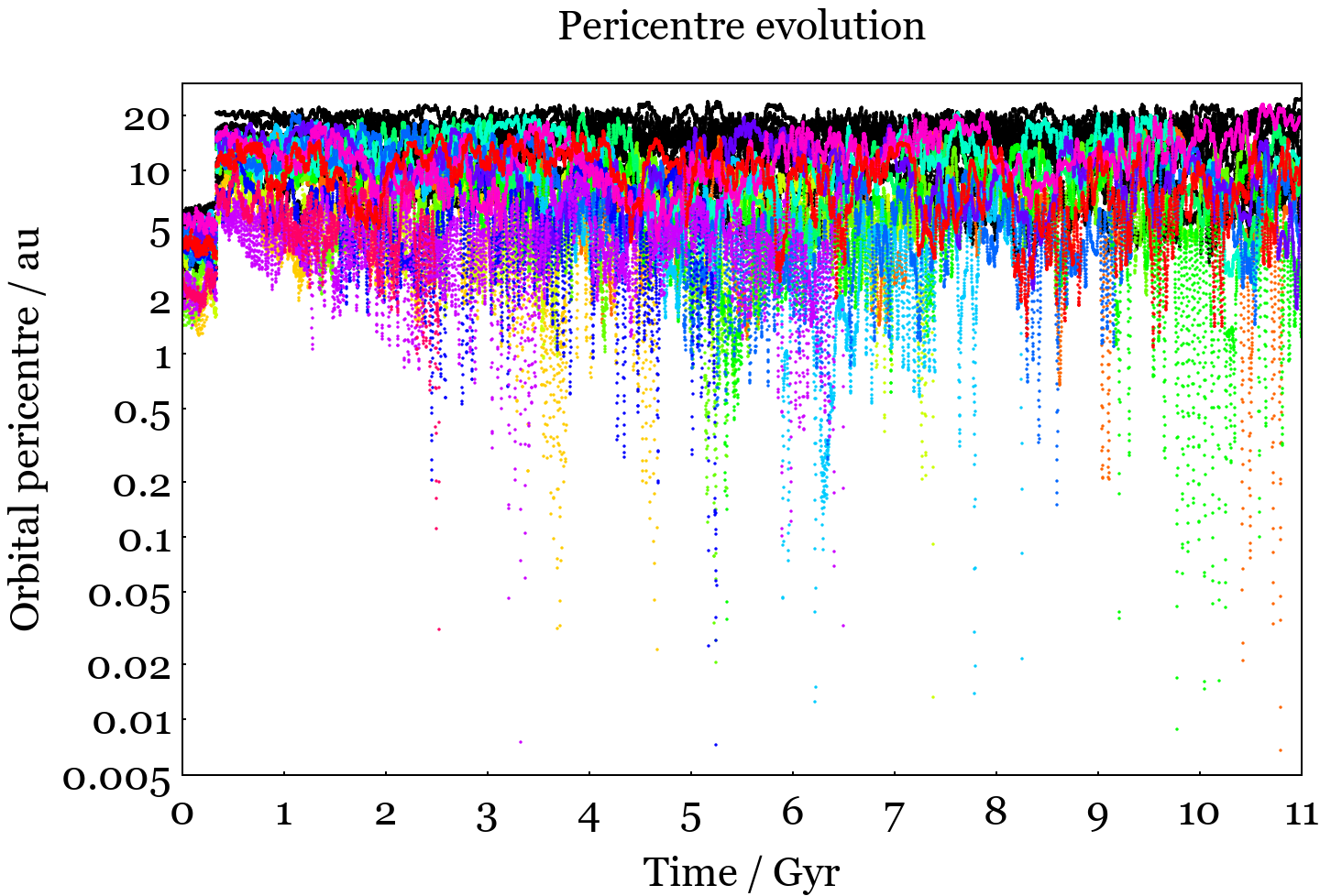}
\includegraphics[width=8.5cm]{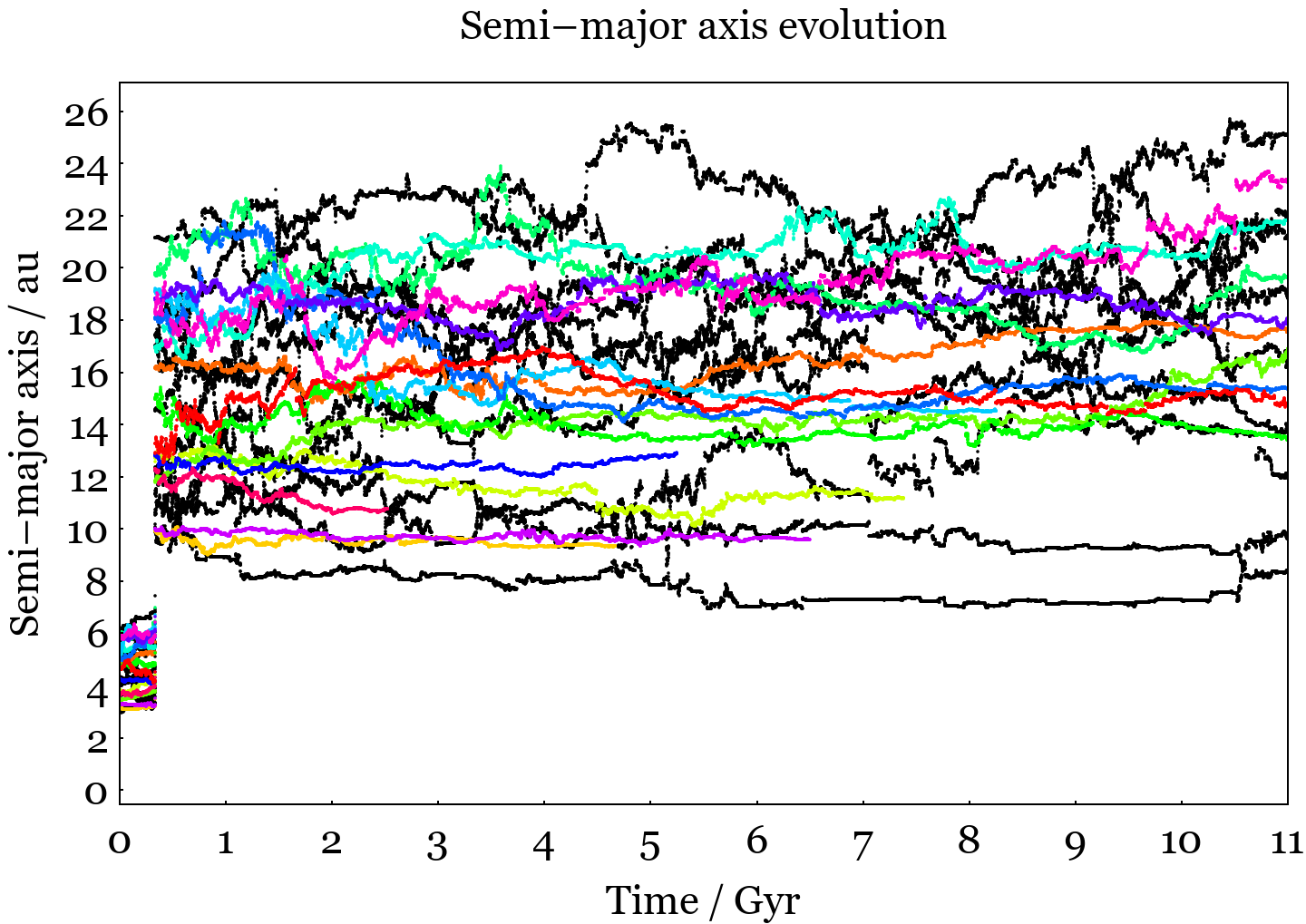}
}
\centerline{}
\centerline{
\includegraphics[width=8.5cm]{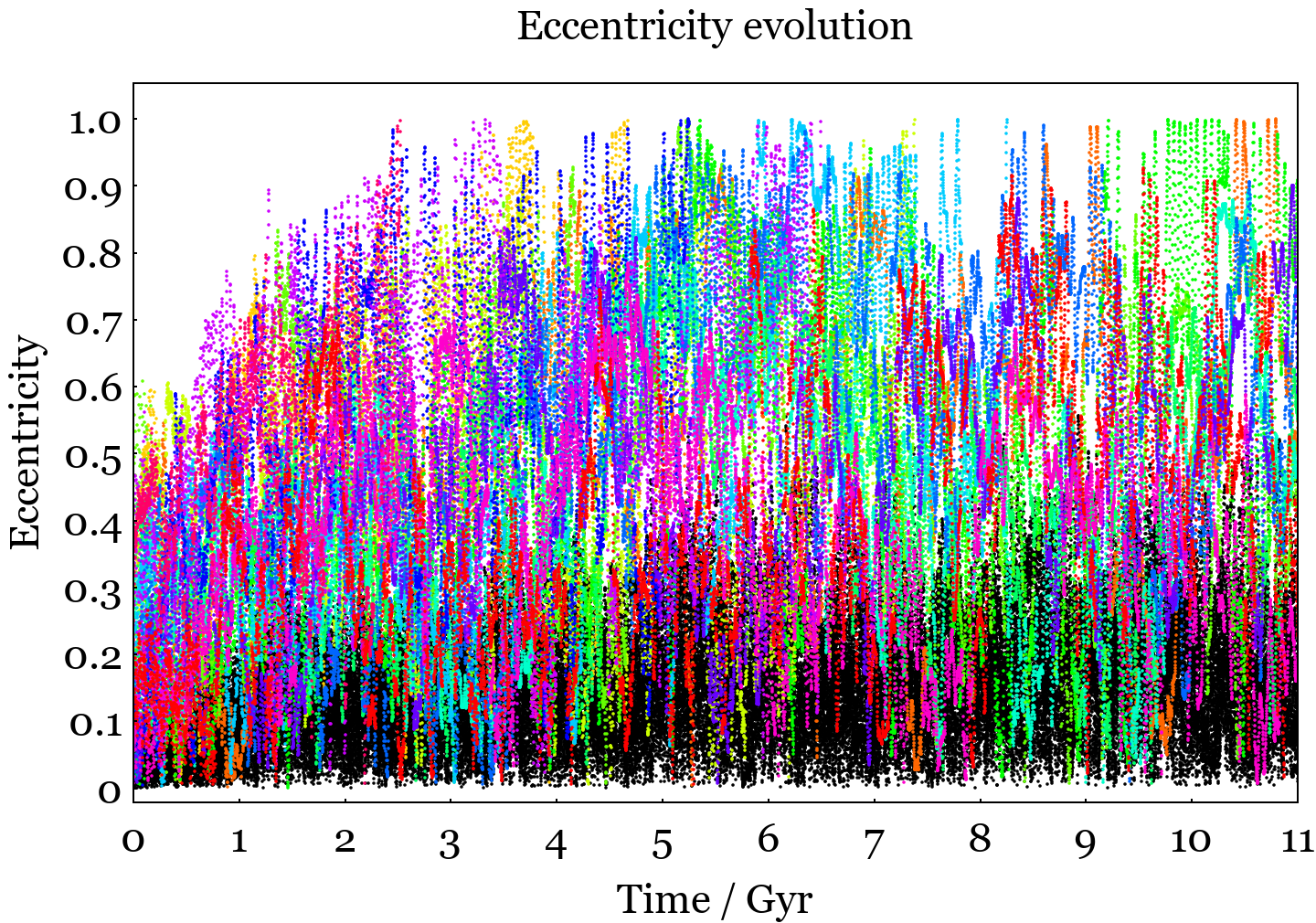}
\includegraphics[width=8.5cm]{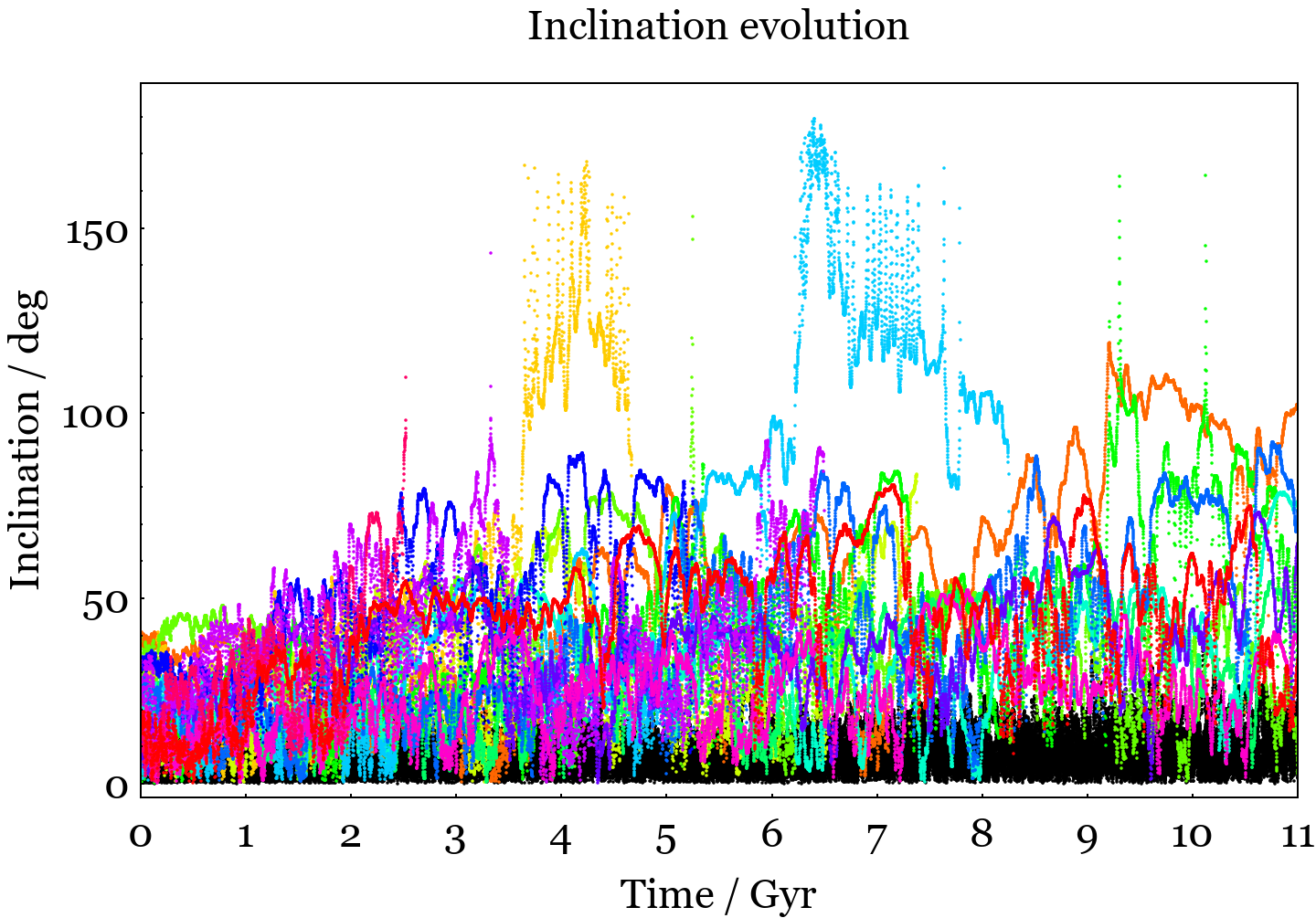}
}
\caption{
Parameter evolutions for fiducial simulations. Shown are the evolutions of the pericentre distance (upper left), semi-major axis (upper right), eccentricity (lower left), and inclination (lower right) for 10 perturbers (black curves) and 15 test particles (coloured curves). The mass of each perturber is $4 M_{\leftmoon}$, and the system is integrated for 11 Gyr from the start of the giant branch phases of a host star with a mass of $2.00M_{\odot}$ on the main sequence. 6 of the 15 test particles are accreted by the white dwarf at cooling ages ranging from 2.2 Gyr to 7.9 Gyr, and other test particles experience significant radial incursions later. These plots demonstrate the capability of sub-terrestrial perturbers, which are somewhere in the mass range $1-10 M_{\leftmoon}$, to pollute white dwarfs.
}
\label{fiducial}
\end{figure*}

\begin{figure*}
\includegraphics[width=16.0cm]{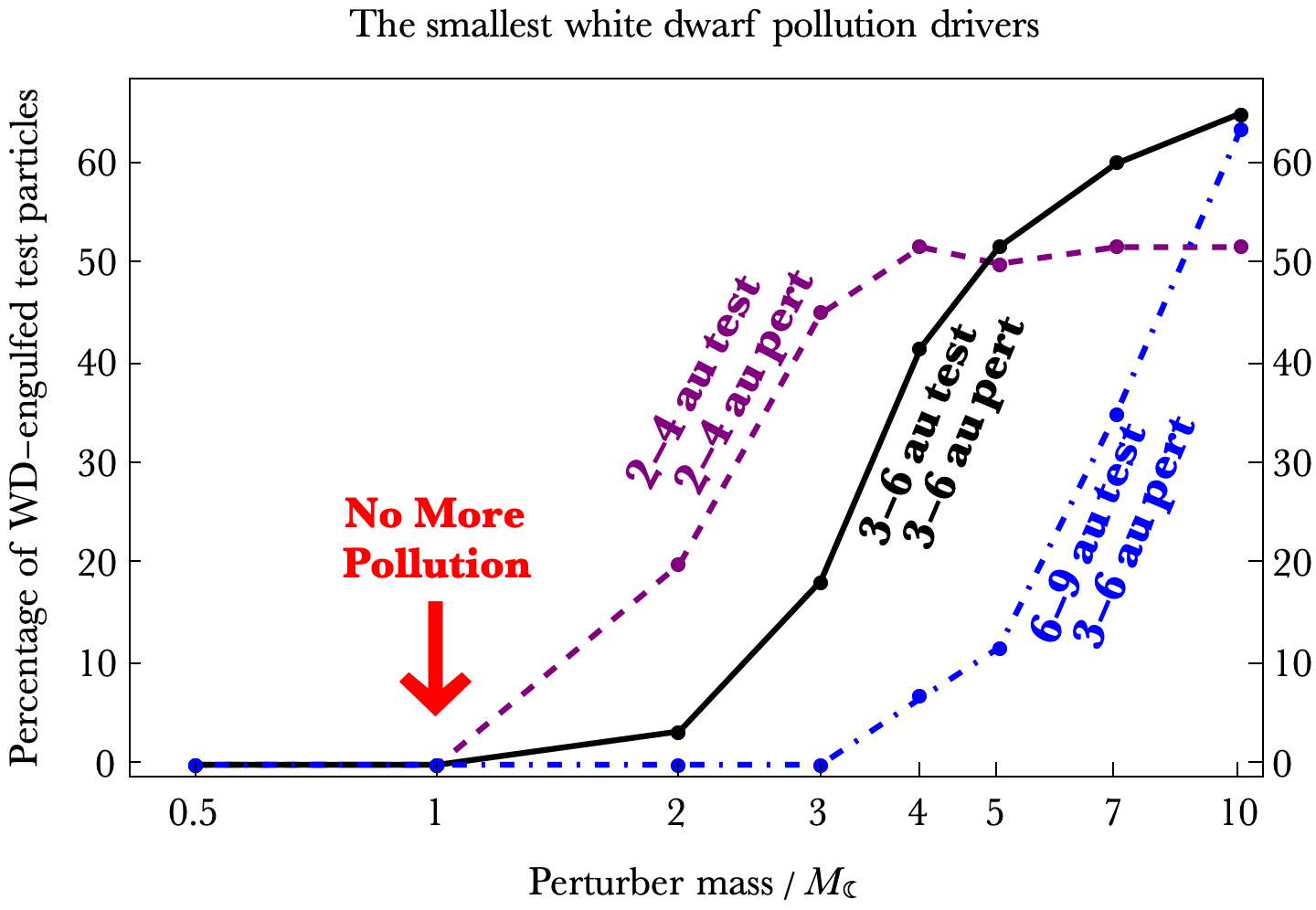}
\caption{
The punchline plot of this investigation, illustrating that the lowest perturber mass that can feasibly pollute white dwarfs is about $1M_{\leftmoon}$. Each dot represents aggregate statistics from a series of 4 simulations with the perturber mass on the $x$-axes and the initial semi-major axis range of {\rev both the test particles and the perturbers, which are labelled on the differently styled and coloured curves.} 
}
\label{punch}
\end{figure*}

\begin{figure}
\includegraphics[width=8.0cm]{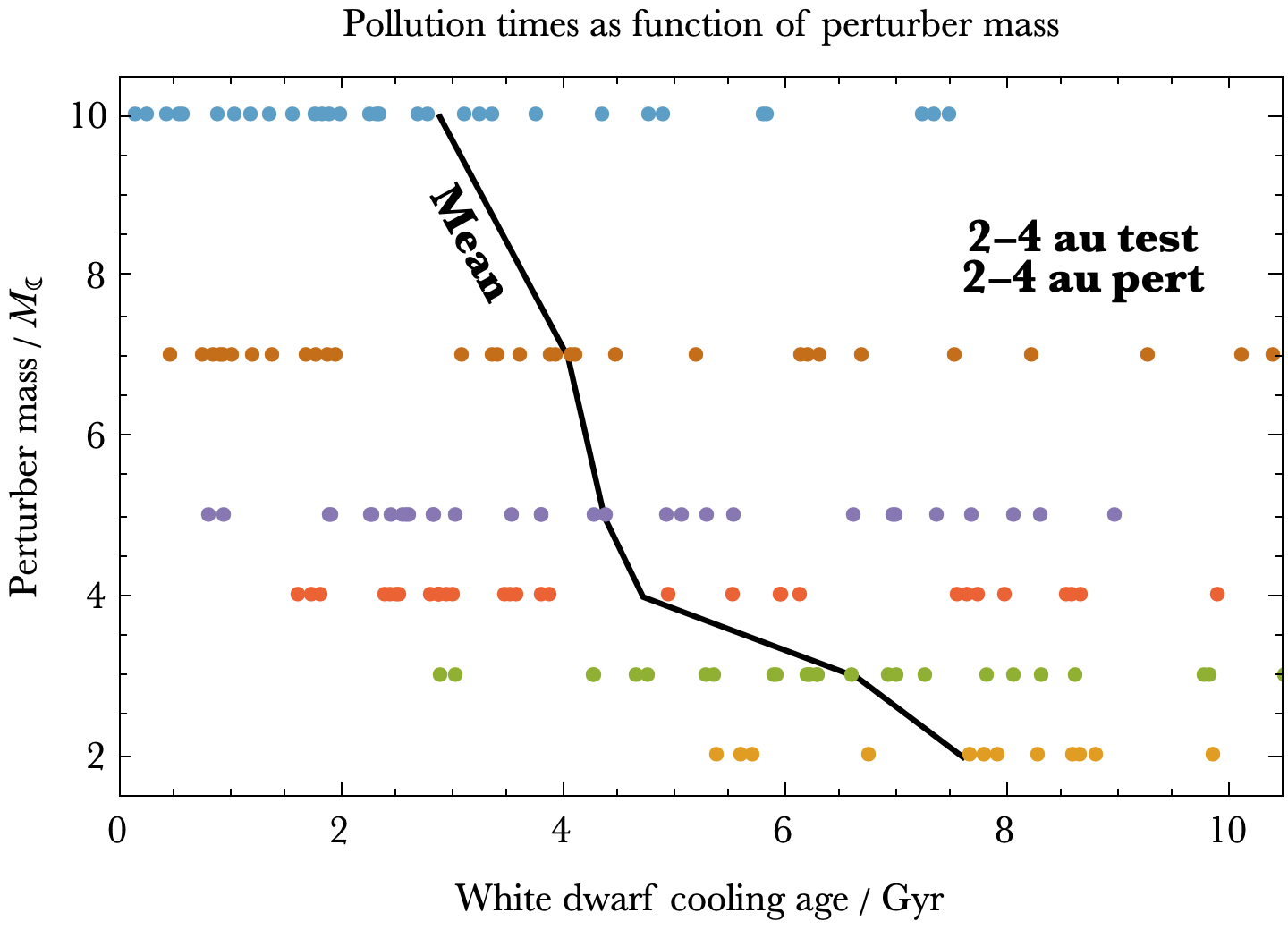}
\centerline{}
\includegraphics[width=8.0cm]{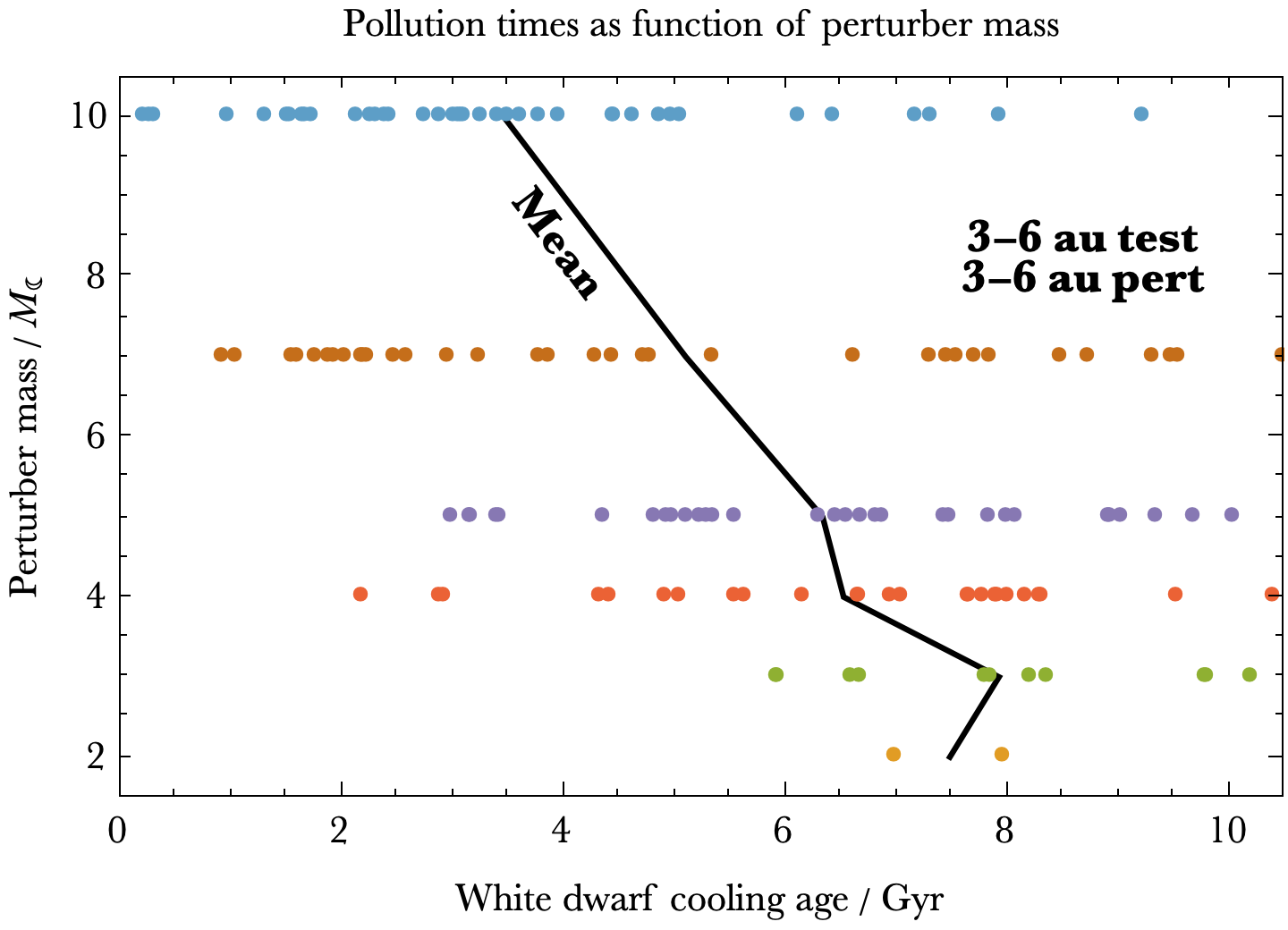}
\centerline{}
\includegraphics[width=8.0cm]{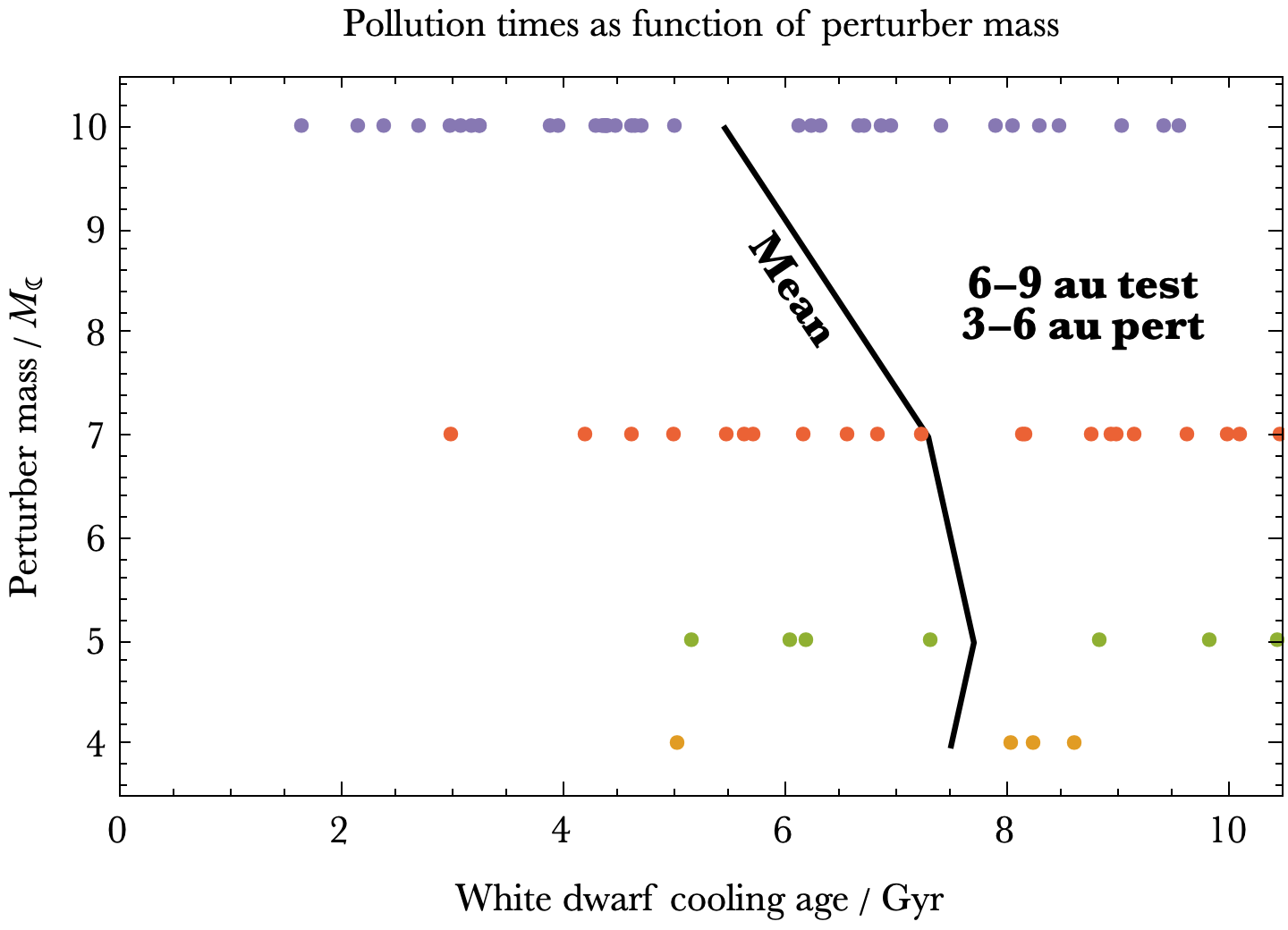}
\caption{
{\rev The distribution of pollution times as a function of perturber mass. The initial semi-major axis range of both the test particles and the perturbers are labelled on each plot. The mean pollution times are connected by solid lines, and demonstrate that despite the wide scatter in accretion events generated by large perturbers, smaller perturbers pollute white dwarfs at later ages.}
}
\label{cool}
\end{figure}

\subsection{Results}

\subsubsection{Figure \ref{fiducial}: A fiducial simulation}

We begin with an example of the evolution of the orbital parameters of one of our fiducial systems {\rev for illustrative purposes}. Figure \ref{fiducial} {\rev shows} the evolution of both the perturbers, in black, and the test particles, in other colours, for $4M_{\leftmoon}$ perturbers.

In this particular system, the perturbers by themselves do become unstable, with two of them colliding with each other at 1.2 Myr (during the giant branch phases). However, for the remainder of the simulation, all other surviving perturbers remain on stable orbits. Their orbits do however vary significantly, as is perhaps most evident in the semi-major axis evolution plot. This plot also reveals that the test particle semi-major axes do not significantly vary as frequently as their eccentricities and inclinations. Their eccentricities can easily reach values near unity after the star has become a white dwarf. Hence, the close pericentre approaches to the white dwarf are primarily driven by changes in eccentricity, not semi-major axis.

Of the 15 test particles, 6 pollute the white dwarf, and do so at cooling ages (which is the time since the star became a white dwarf) between 2.2 Gyr and 7.9 Gyr. All other 9 particles remain stable, although the plots indicate that two more were almost polluted at cooling ages beyond about 9 Gyr, when their eccentricity oscillations were maximised. 

Overall the figure demonstrates that $4M_{\leftmoon}$ is a clearly pollutable perturbing mass. This demonstration prompted us to perform suites of simulations in order to pinpoint how much we can lower this mass and still generate pollution.

\subsubsection{Figure \ref{punch}: The punchline plot}

{\rev In order to determine the limiting perturber mass}, we performed a total of {\rev 96} simulations. We broke down these {\rev 96} simulations into 16 different parameter groups of perturber mass and test particle semi-major axis range. For each of these 16 pairs of parameters, we performed 4 simulations in order to buttress our statistics.

We report {\rev our} results in Fig. \ref{punch}, which showcases the main result of our investigation. Displayed is the per cent of the test particles that were accreted by the white dwarf as a function of perturber mass and initial test particle semi-major axis. The figure illustrates that the minimum perturber mass can be well approximated by $1-2M_{\leftmoon}$, {\rev even if the innermost main-sequence semi-major axis of the debris and perturbers are set at just 2 au.} 

The curves in this plot also demonstrate that (i) at the higher end of the perturber masses that we sampled, over half of the test particles were accreted, and (ii) perturbers can efficiently accrete external debris as long as the perturber mass is large enough. The latter aspect has been known for over a decade \citep{bonetal2011}, but not necessarily with such low perturber masses.

\subsubsection{{\rev Figure \ref{cool}: Correlation with cooling age}}

{\rev These 96 simulations also reveal an important connection between perturber mass and cooling age at which the white dwarf is polluted. We illustrate this connection {\rev in} Fig. \ref{cool}.}

{\rev The figure displays all instances where the white dwarf was polluted (the giant branch phase is not shown). For most perturber masses, there is a wide scatter in pollution times. Nevertheless, an overall trend is apparent for each set of initial semi-major axis choices: the solid lines connect the mean pollution times at each perturber mass, and illustrate that on average older white dwarfs become polluted by smaller mass perturbers.} This trend is perhaps expected, but has been quantified previously only for much higher perturber masses \citep{frehan2014,musetal2018,veretal2021,ocoetal2022}.

{\rev A secondary conclusion from these plots is that the smallest perturbers cannot seem to pollute the white dwarf at young cooling ages. In fact, for $M_{\rm per} = 2 M_{\leftmoon}$, no pollution occurs in any simulation for cooling ages under 5 Gyr. In contrast, the most massive perturbers investigated here may pollute the white dwarf at any cooling age. These observations, however, should be contextualized with our small number statistics.}

\begin{figure*}
\centerline{\Huge \underline{Variation of perturber mass and test particle distance}}
\centerline{}
\centerline{}
\centerline{\Huge \ \ \ \ \ \ \ 3-6 au on main-sequence  \   
                                \ \ \ \   6-9 au on main-sequence}
\centerline{}
\centerline{
\includegraphics[width=8.5cm]{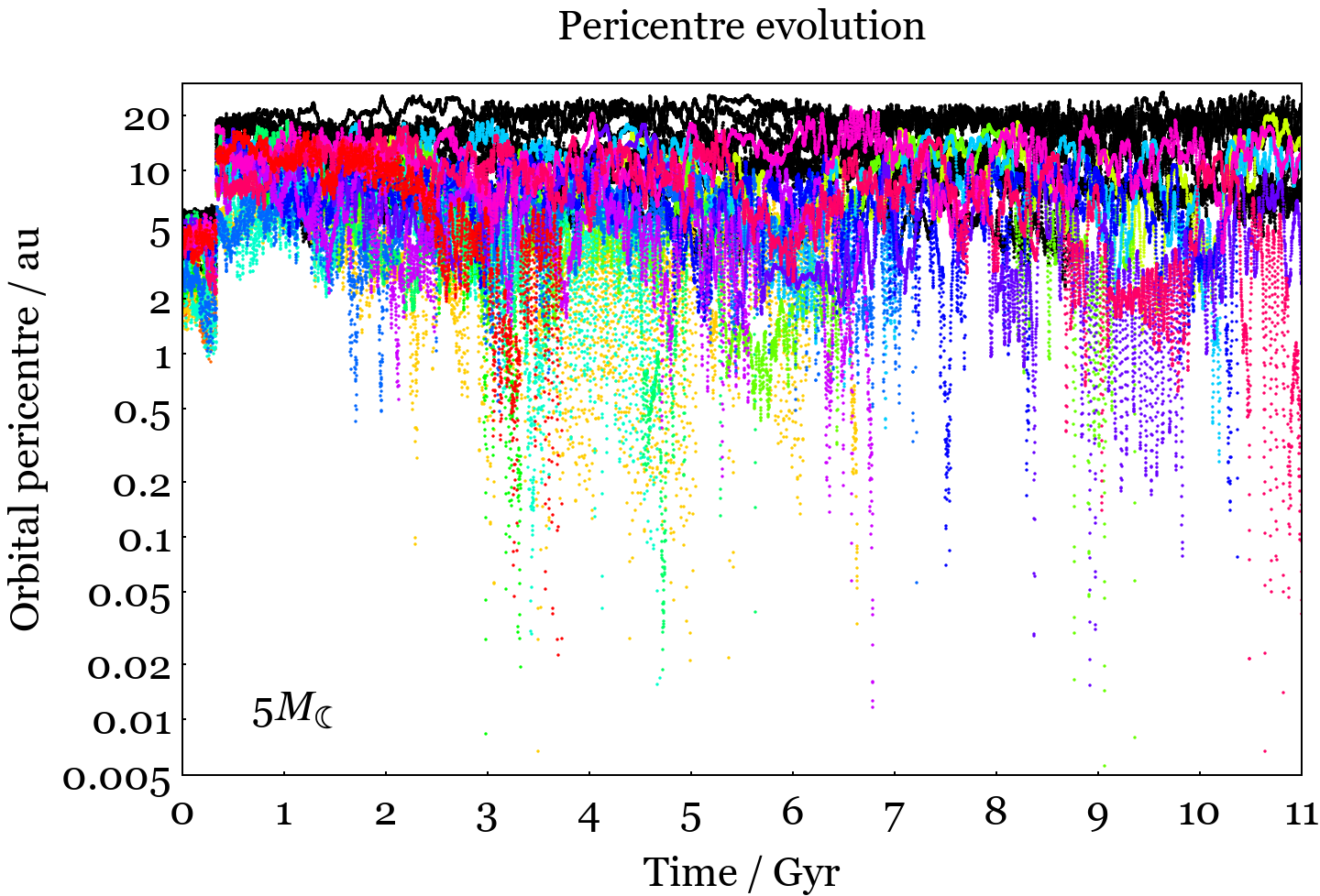}
\includegraphics[width=8.5cm]{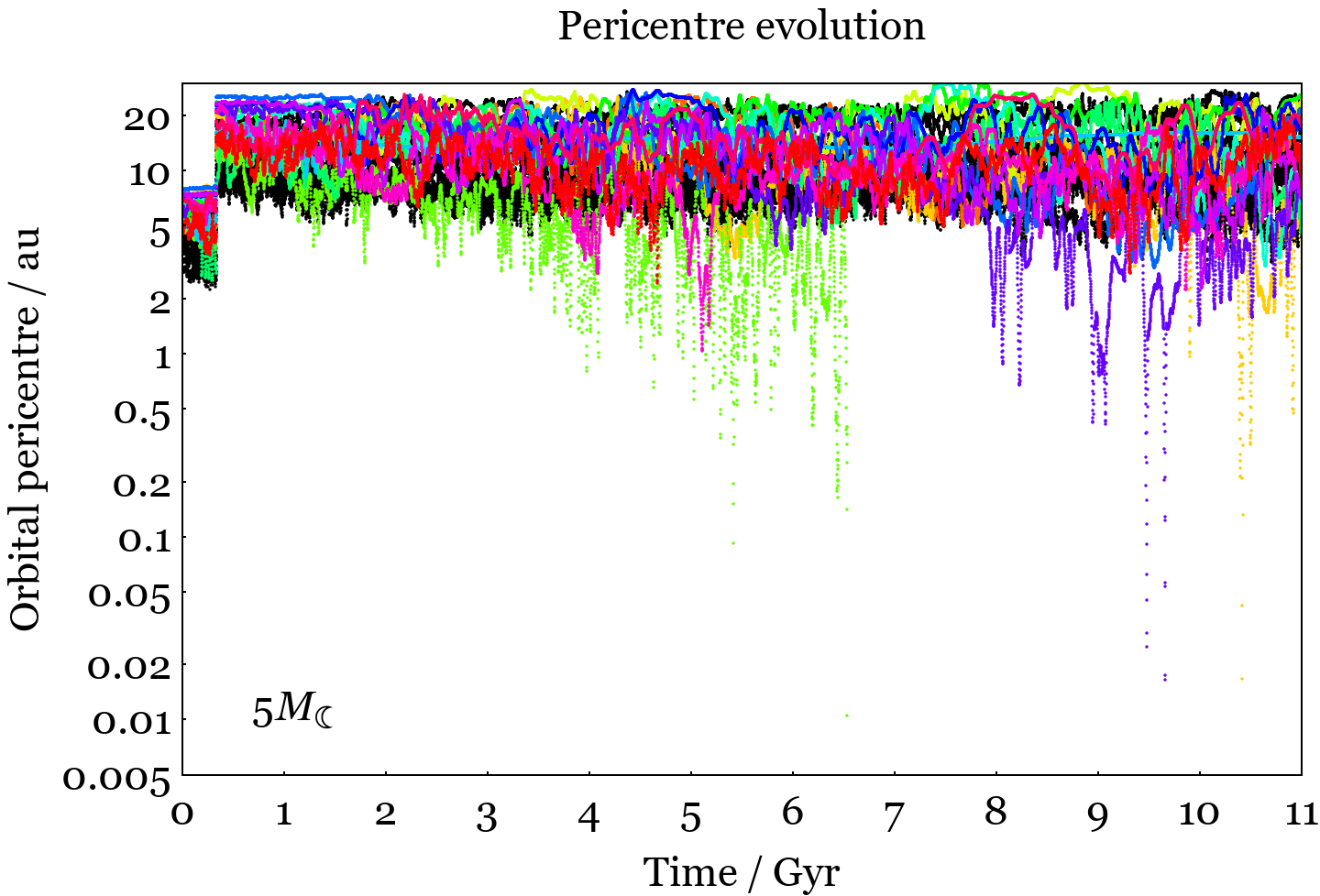}
}
\centerline{}
\centerline{
\includegraphics[width=8.5cm]{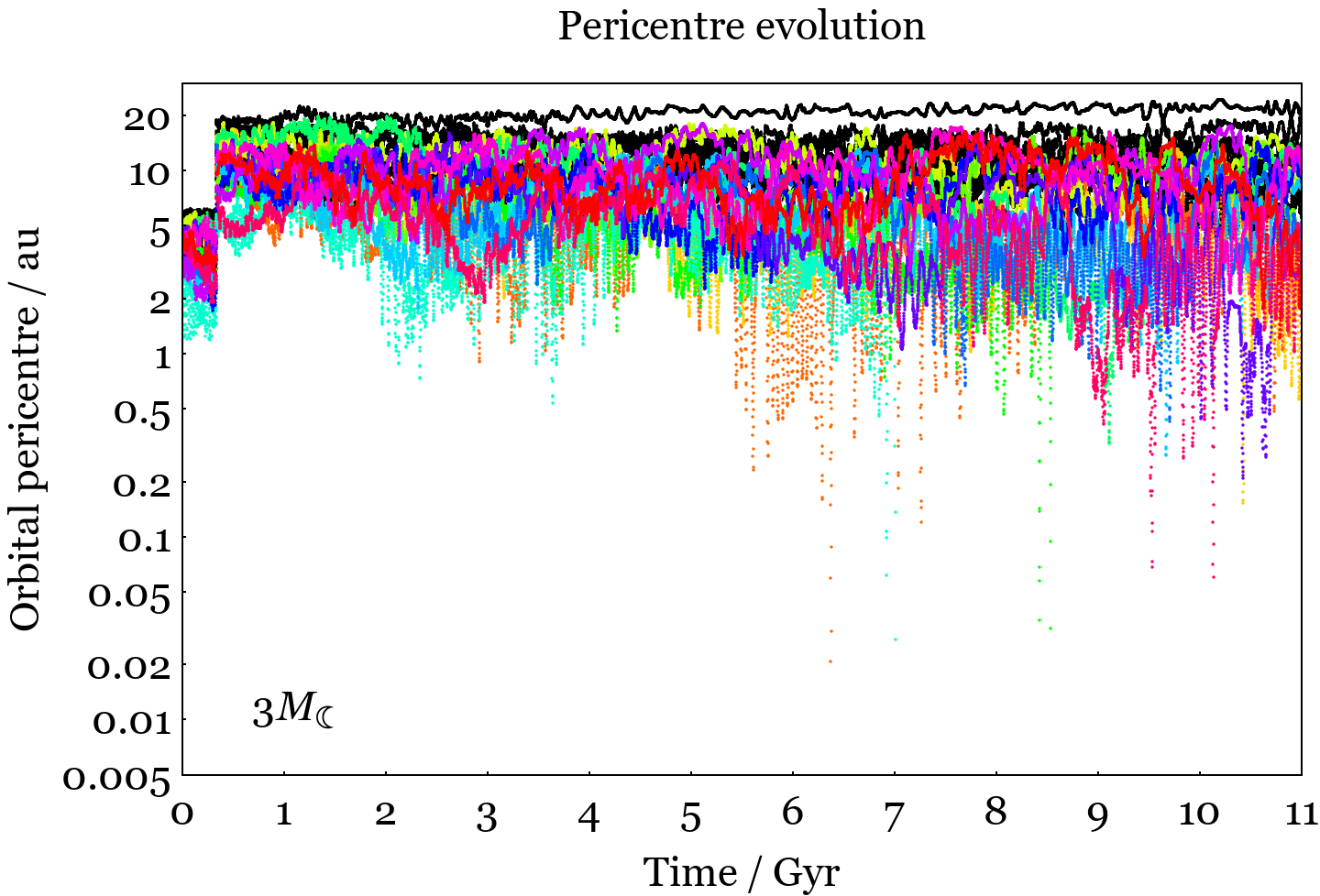}
\includegraphics[width=8.5cm]{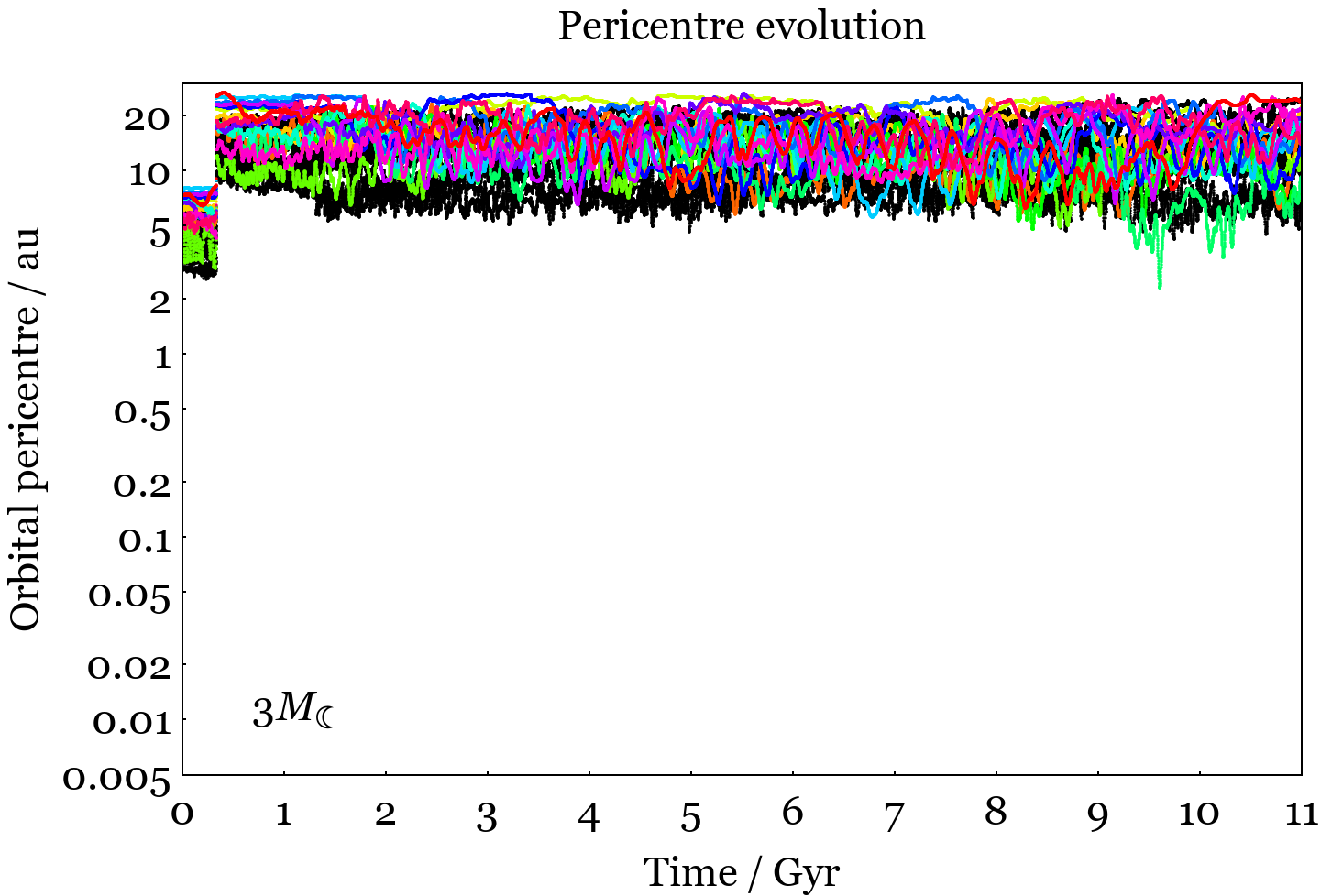}
}
\centerline{}
\centerline{
\includegraphics[width=8.5cm]{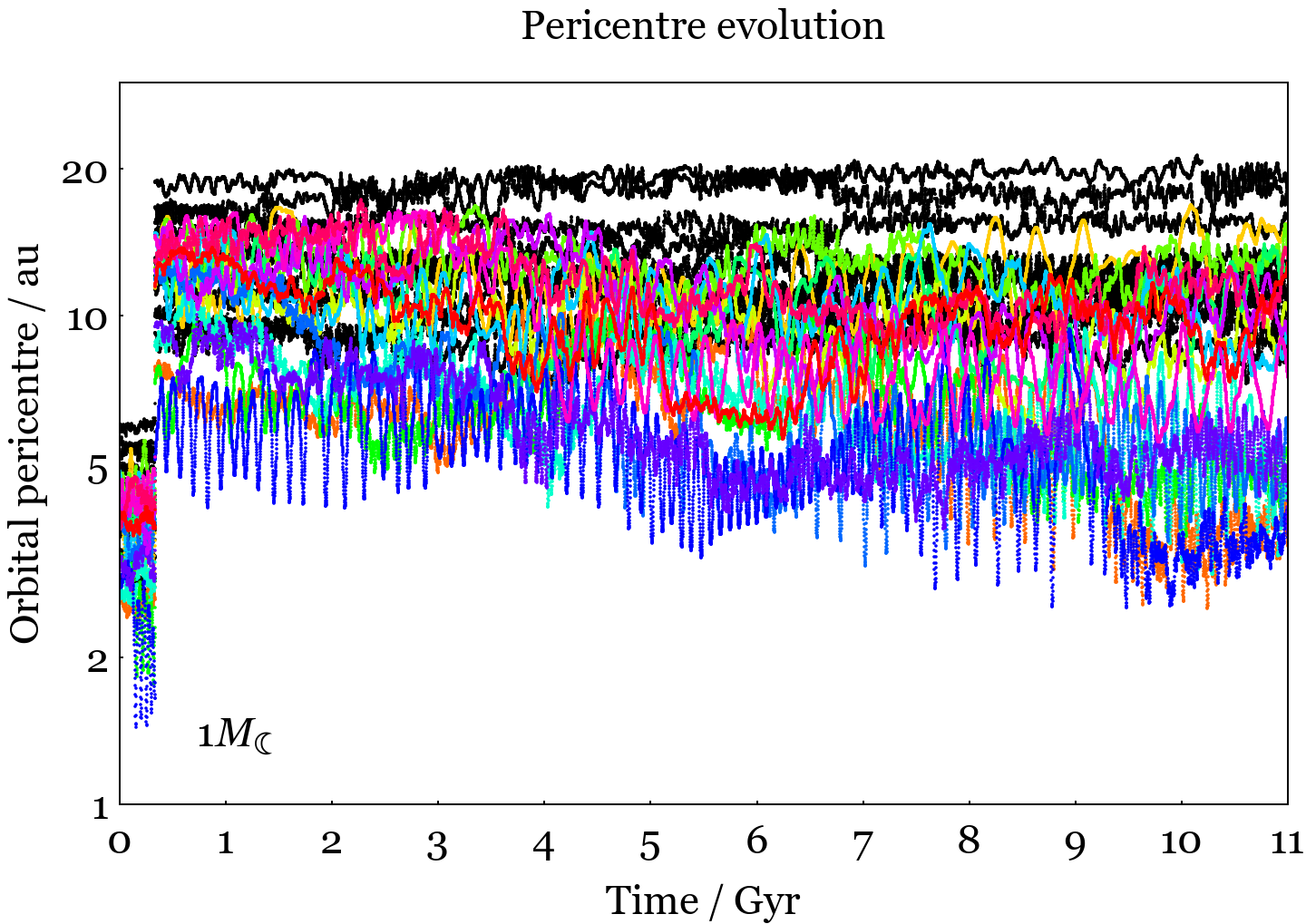}
\includegraphics[width=8.5cm]{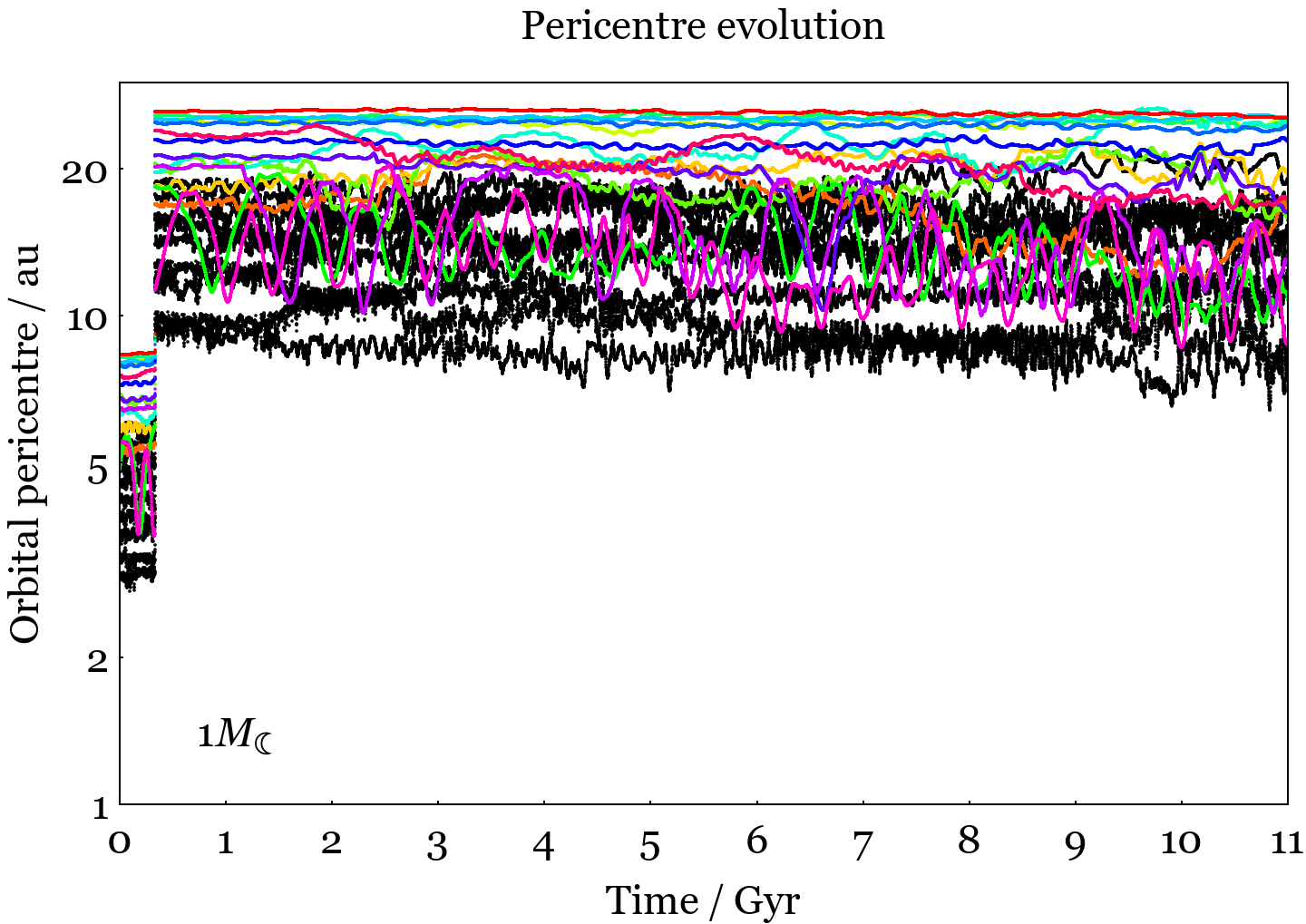}
}
\caption{
How the pericentre evolution changes as a function of perturbation mass and initial test particle semi-major axis. Perturbers are represented by black curves, and test particles are represented by coloured curves. The perturber masses in the simulations shown are $5M_{\leftmoon}$ (top row), $3M_{\leftmoon}$ (middle row) and $1M_{\leftmoon}$ (bottom row), and the initial semi-major axis ranges for the test particles are $3-6$ au (left column) and $6-9$ au (right column). Pollution onto the white dwarf occurs only in the top row (both plots) and the middle left plot. Hence, the vertical scale of the plots in the lower row is different from the vertical scale in the other plots, and this scale difference helps to illustrate the extent of the perturbations when the initial semi-major axis values are varied. This figure illustrates the same result as in Fig. \ref{punch}, except with more explicit simulation detail.
}
\label{distance}
\end{figure*}

\subsubsection{Figure \ref{distance}: Trends through individual system evolutions}

In order to show in more explicit detail the trends illustrated in {\rev Figs. \ref{punch}$-$\ref{cool}}, we present the evolution of the pericentres for 6 of the simulations in Fig. \ref{distance}. In this figure, the mass of the perturbers is varied from top to bottom from $5M_{\leftmoon}$ to $3M_{\leftmoon}$ to $1M_{\leftmoon}$, and the vertical scale in the bottom row plots is shifted because no pollution occurred for those simulations. By comparing the left and right columns, one can contrast how external test particles are not excited as much as those that share similar semi-major axis space as the perturbers. These plots also illustrate how decreasing the perturber mass by a factor of just a few can prevent test particles from ever reaching distances within about 1 au.

\begin{figure*}
\centerline{\Huge \underline{Variation of parameters from fiducial cases}}
\centerline{}
\centerline{\Huge \ \ \ \ \  Changing stellar mass  \   
                             \ \ \ \ \ \    Changing \# of perturbers}
\centerline{}
\centerline{
\includegraphics[width=8.5cm]{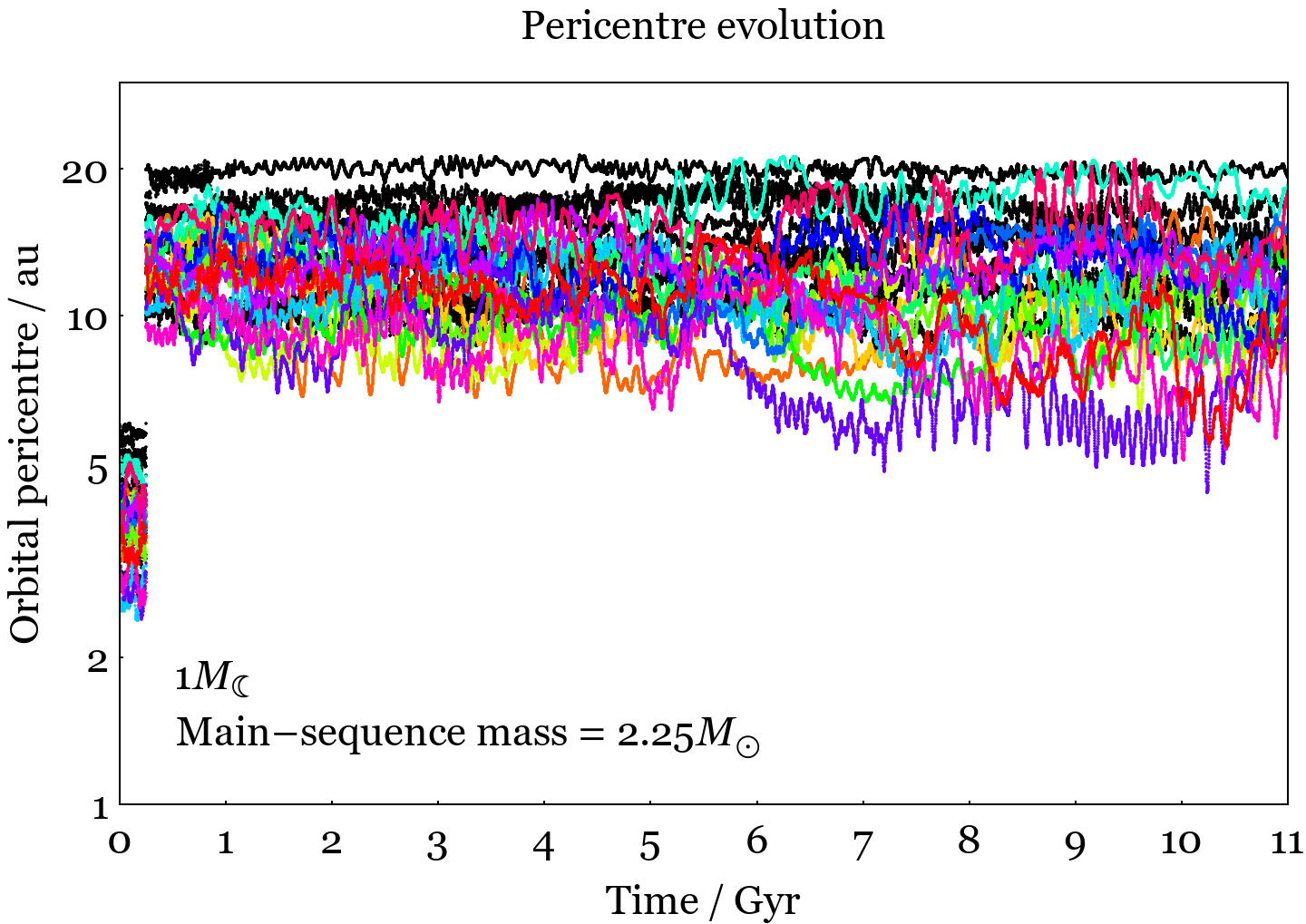}
\includegraphics[width=8.5cm]{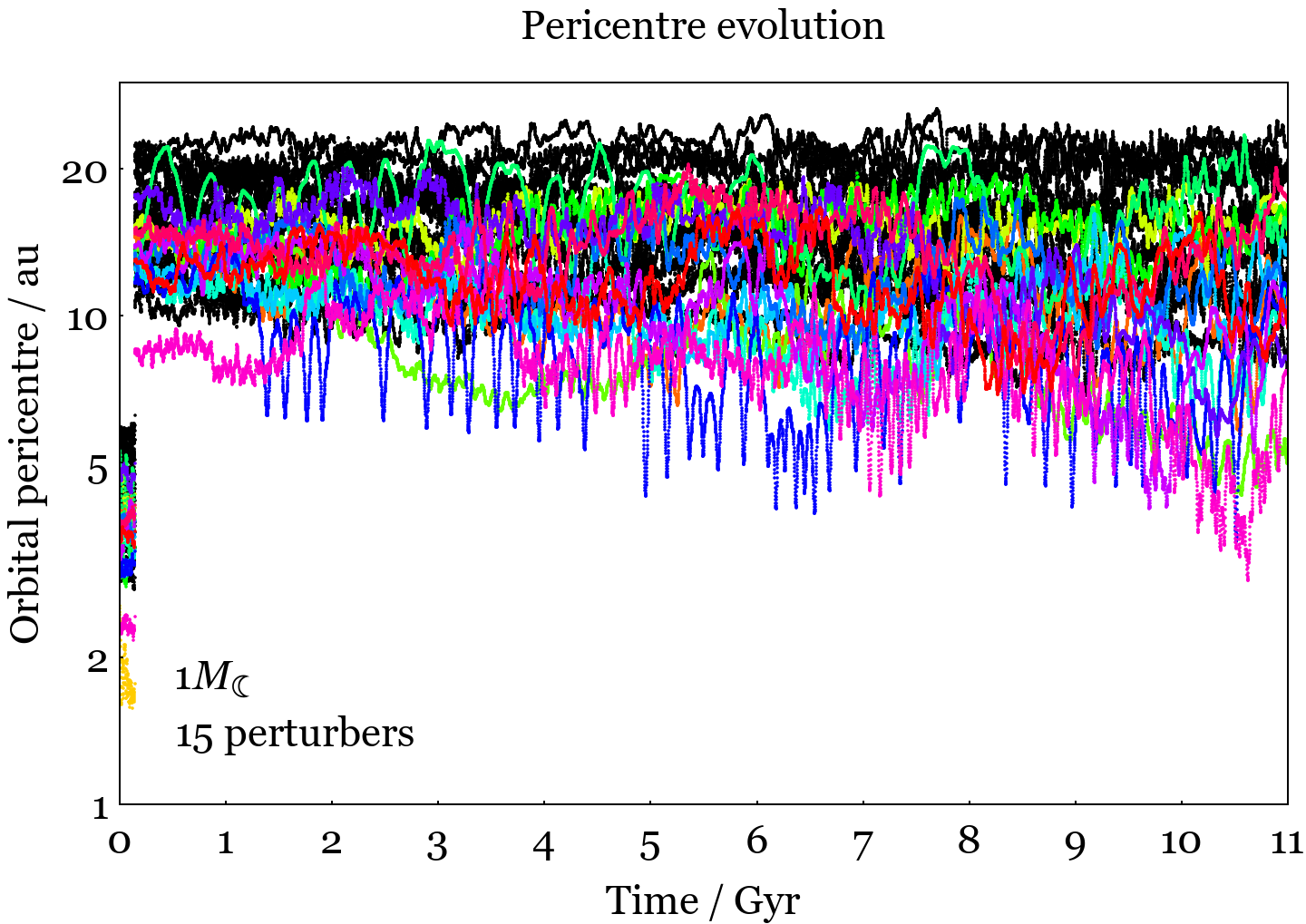}
}
\centerline{}
\centerline{
\includegraphics[width=8.5cm]{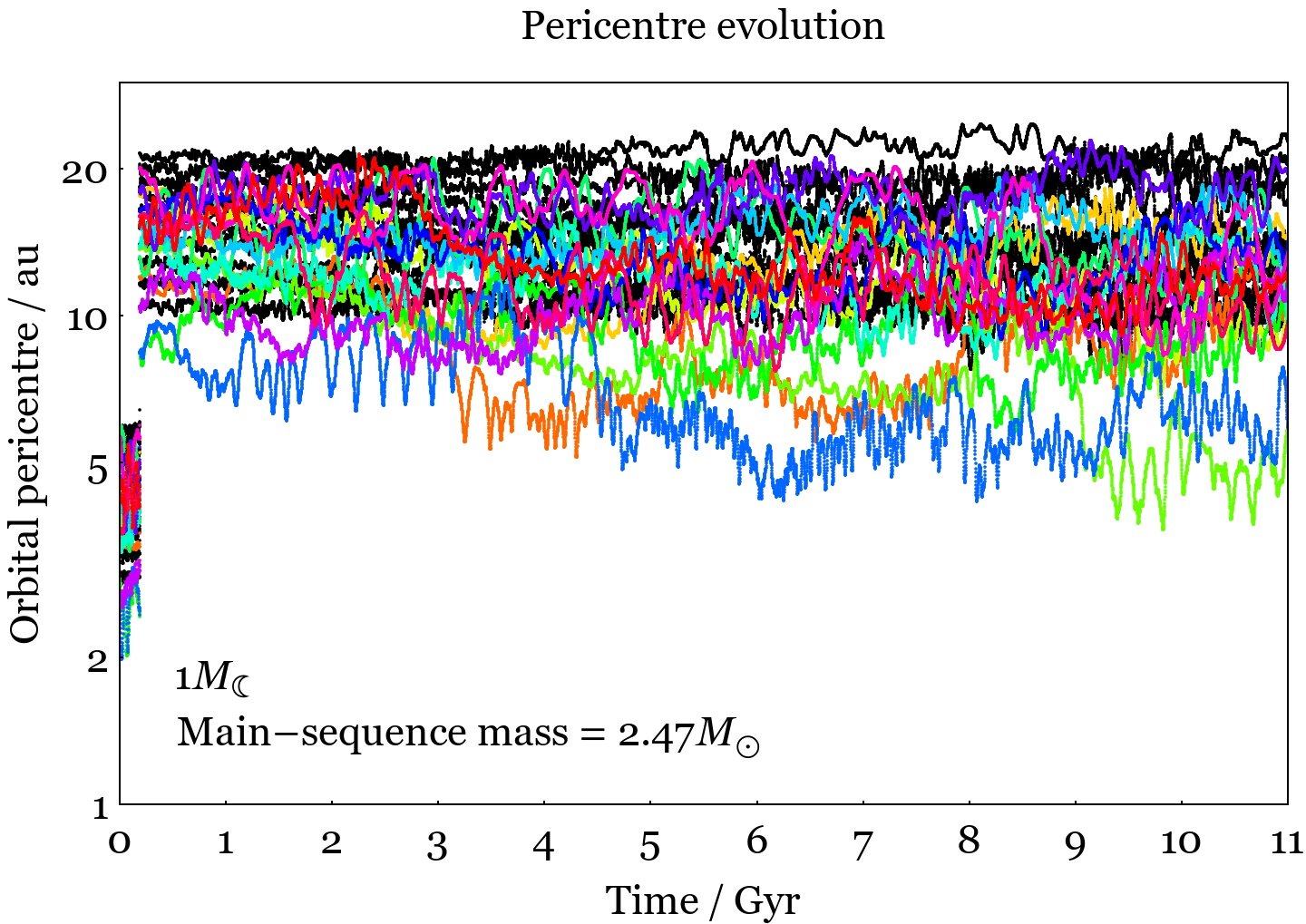}
\includegraphics[width=8.5cm]{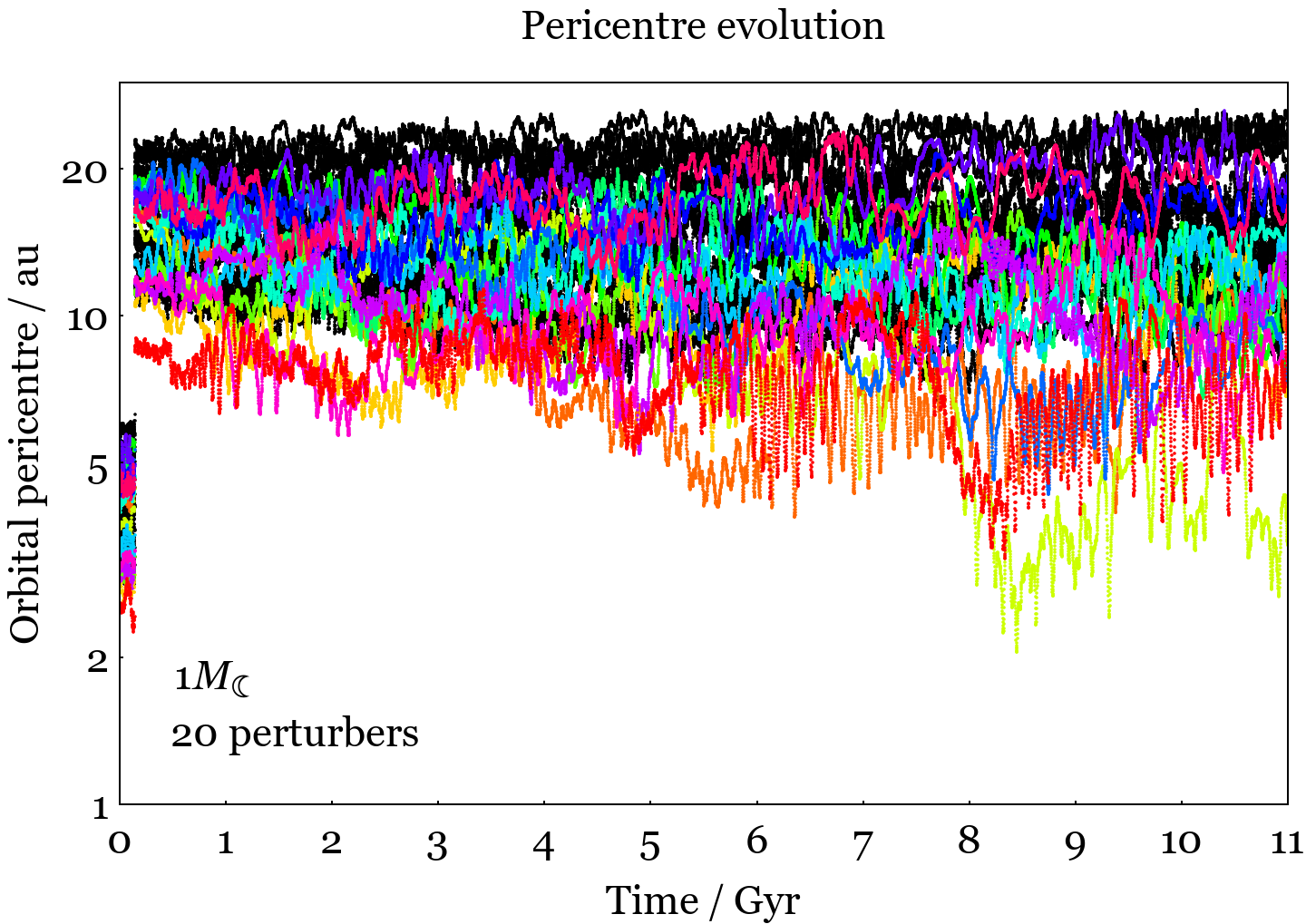}
}
\centerline{}
\centerline{
\includegraphics[width=8.5cm]{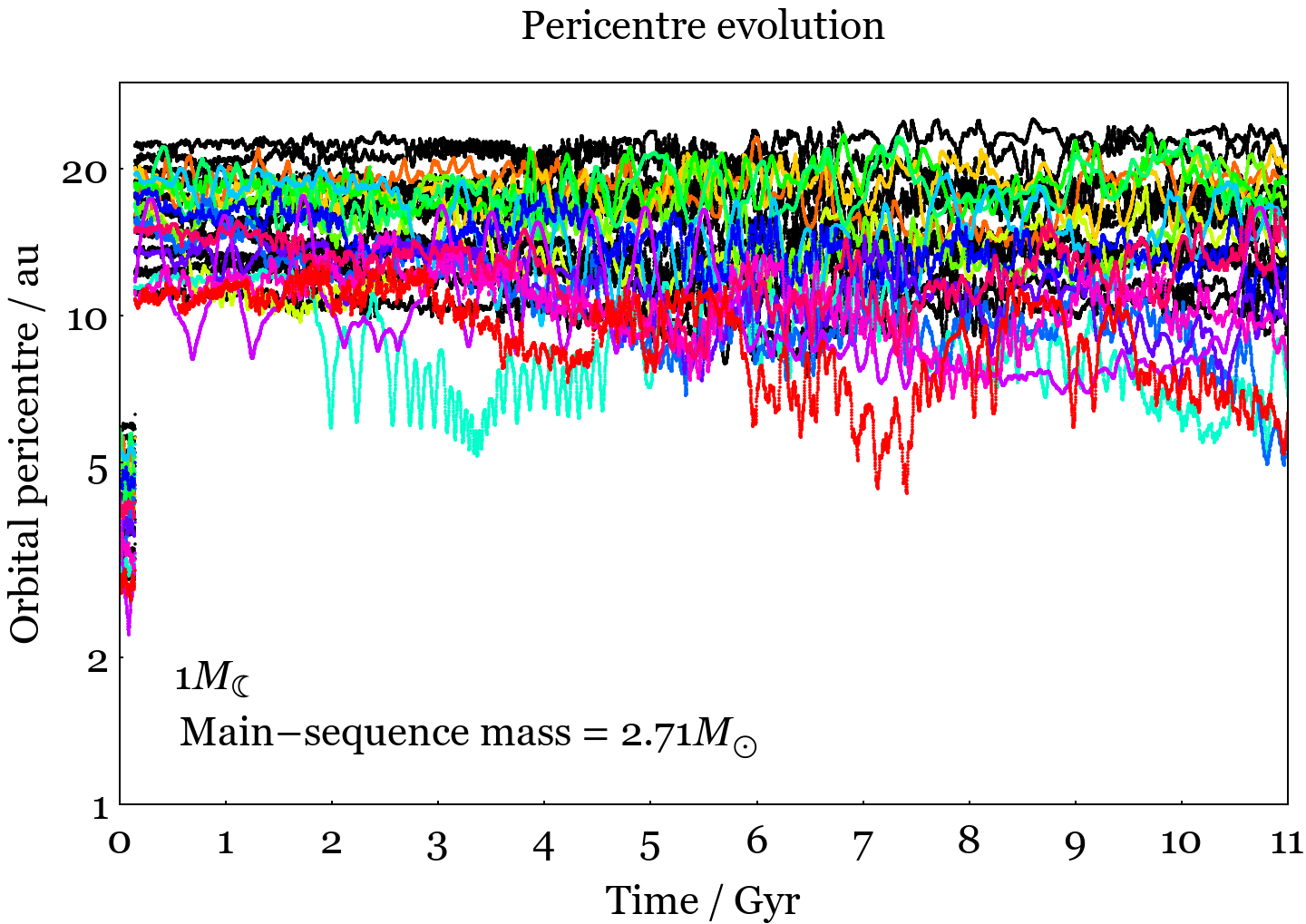}
\includegraphics[width=8.5cm]{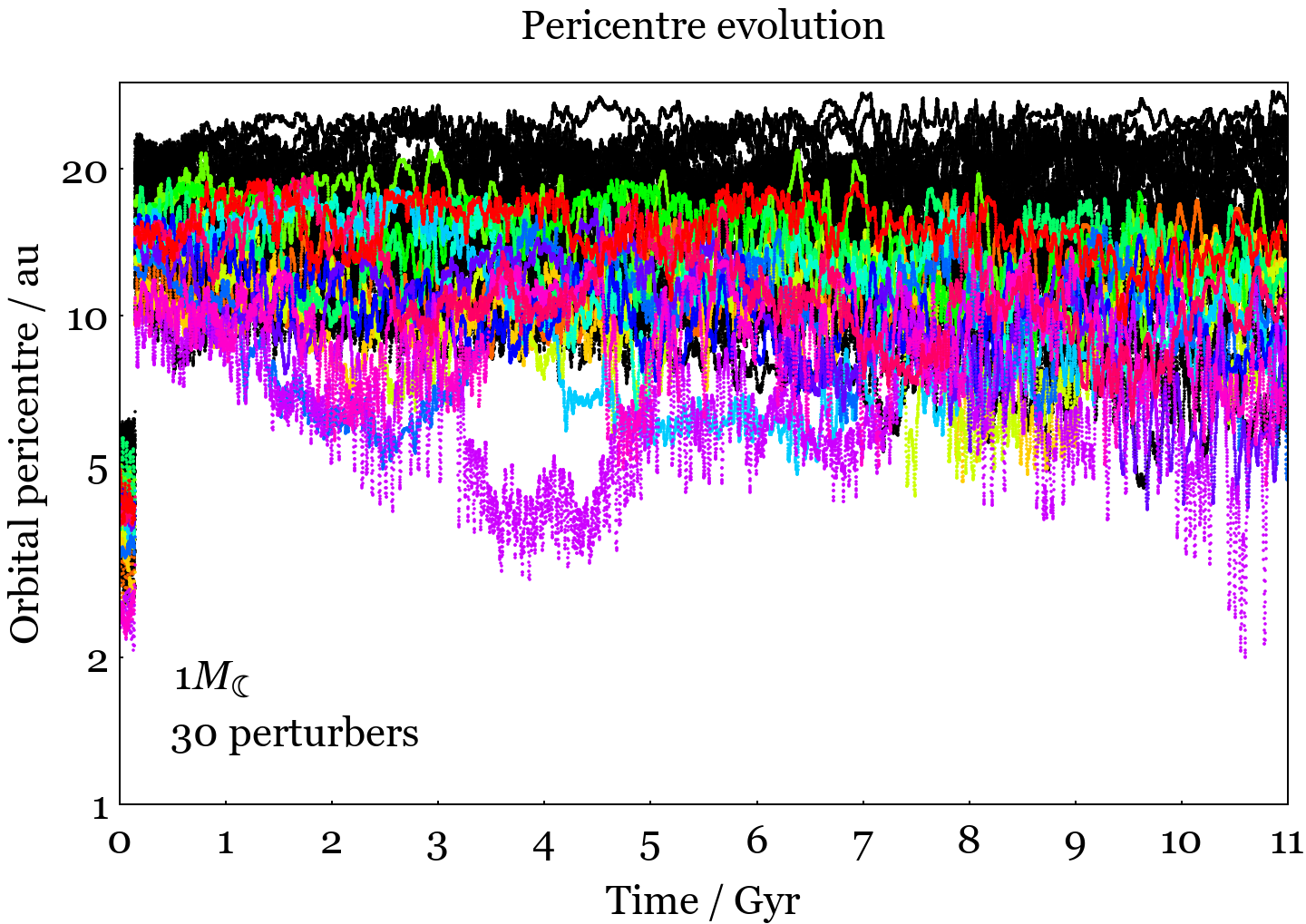}
}
\caption{
How the pericentre evolution changes as the stellar mass and number of perturbers are varied. The specific values are provided in the inset text. The stellar mass is increased as one moves down the left column of the figure, and the number of perturbers are increased as one moves down the right column of the figure. In all plots, the perturber mass is $1M_{\leftmoon}$, and none of the white dwarfs in any of the plots are polluted by test particles. This result helps strenghen the notion that $1M_{\leftmoon}$ is a feasible lower limit for a perturber mass to generate pollution.
}
\label{stellarmass}
\end{figure*}

\begin{figure*}
\includegraphics[width=19.0cm]{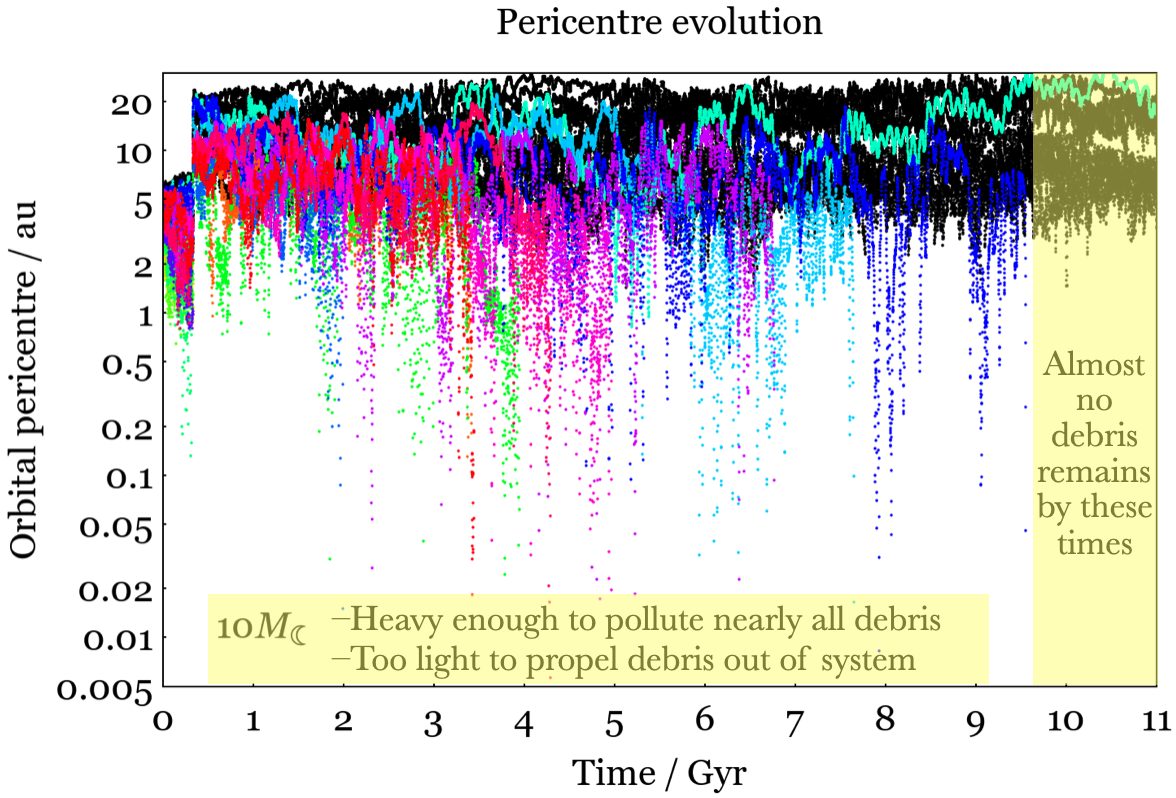}
\caption{
An example of pericentre evolution with the most massive perturber that we modelled ($10M_{\leftmoon}$), which still represents a much lower perturber mass than what is usually found in the literature. This plot illustrates that $10M_{\leftmoon}$ is large enough to clear the system of debris by the oldest cooling ages, but small enough such that the clearance is almost entirely in the form of pollution onto the white dwarf rather than escape from the system. Hence, objects with masses on the order of $10 M_{\leftmoon}$ may be efficient but unheralded polluters.
}
\label{massive}
\end{figure*}

\subsubsection{Figure \ref{stellarmass}: Robustness of result}

How robust is the result presented in Fig. \ref{punch}? In order to explore this question, we have performed additional simulations where we have varied the initial stellar mass and total number of perturbers (while keeping the 3-6 au semi-major axis range fixed). We present the results in Fig. \ref{stellarmass}.

In some of these extra simulations, we have evolved host stars with main-sequence masses and giant branch durations of $2.25M_{\odot}$/1.08 Gyr, $2.47M_{\odot}$/0.829 Gyr and $2.71M_{\odot}$/0.638 Gyr. These three types of stars lose a higher percentage of their mass, and more quickly, than $2.00M_{\odot}$ stars, creating more dynamically active environments and increasing the prospects for instability. Nevertheless, as the pollution-less examples on the left column of the figure show, this increase in stellar mass does not appear to have a strong effect on the minimum pollutable perturber mass.

In some other simulations, we increased the number of perturbers to 15, 20, and then 30. As the examples in the right column of the figure demonstrate, this increase does lead to more dynamically active test particles, but not necessarily to any more pollution. Note that the lower limit of the $y$-axis scale of each plot is 1 au; none of the particles achieve pericentres inside of this value. 

\section{Discussion} 

Identifying the minimum pollutable perturber mass is useful in order to inform future modelling efforts, particularly when trying to reproduce ever-changing observational constraints on accretion rates as a function of time \citep{koeetal2014,holetal2018,bloxu2022}. A dichotomy currently exists in which we are aware of giant planets orbiting white dwarfs \citep{thoetal1993,sigetal2003,luhetal2011,ganetal2019,vanetal2020,blaetal2021,gaietal2022} and minor planets smaller than Ceres disintegrating around other white dwarfs \citep{vanetal2015,manetal2019,vanderboschetal2020,guietal2021,faretal2022}. However, we have little observational information about orbiting objects with masses in-between these two extremes.

{\rev Surprisingly}, this hidden population could represent the most efficient polluters. Consequently, understanding the polluting capabilities of this population is crucial. Here, although we have focussed on identifying the approximate lowest pollutable perturber mass ($1M_{\leftmoon}$), our higher mass perturber simulations are also revealing. 

Figure \ref{massive} illustrates the pericentre evolution of one of our simulations with $10M_{\leftmoon}$ perturbers. These perturbers are efficient: of the 15 test particles, 5 are engulfed by the star during the giant branch phases, 9 pollute the white dwarf, and only 1 survives until the end of the simulation. Further, the clearence of test particles in this simulation occurs too early to accrete onto the polluted white dwarfs with the greatest known cooling ages \citep{elmetal2022}. In this sense, these perturbers are {\it too massive}, despite being sub-terrestrial.

On the other hand, one key aspect of the perturbers modelled here is that none are sufficiently massive to propel the test particles out of the system. This property is a key reason for their success as polluters: the test particles linger in the system because they cannot escape. Also helpful for pollution prospects is that the perturbers are sufficiently light and small that collisions between the perturbers and test particles are especially unlikely. 

{\rev The test particle approximation of our simulations might begin to break down when the perturber mass is low enough such that it is comparable in mass to the debris disc that the test particles are supposed to represent. The current mass of the solar system Main Belt is just $\approx 0.03 M_{\leftmoon}$, which is significantly lower than the mass of the perturbers considered here. Regardless, extrasolar asteroid belts might be much more massive \citep{debetal2012}.} 

{\rev In cases of comparable disc and perturber mass, feedback on the perturber orbits would need to be modelled. Doing so would significantly increase the computation time of already computationally expensive simulations. In the absence of performing those simulations, we can place disc mass bounds on the validity of our result for the lowest perturber mass. Given that the lowest perturbing mass is about $1M_{\leftmoon}$, then total disc masses of at least one order-of-magnitude lower ($\lesssim 0.1M_{\leftmoon}$) would likely satisfy the assumptions in our simulations. However, we can not rule out that a higher disc mass could decrease the lowest perturbing mass even further.}

Finally, we note that perturbers in the mass range $1M_{\leftmoon} - 1M_{\oplus}$ can form primordially {\rev around} stars, or represent stripped moons of super-Earths or giant planets \citep{payetal2016,payetal2017,trietal2022}. Also, small debris, here represented as test particles, are likely much more mobile than the perturbers during the giant branch phases due to the radiative Yarkovsky effect \citep{veretal2015a,veretal2019}. Further, the debris may be generated freshly at any point during the giant branch phases due to YORP breakup of boulders or asteroids \citep{veretal2014b,versch2020}. Partly for these reasons have we randomly selected semi-major axes for our test particles.

\section{Summary}

Understanding the range of perturber masses that can pollute white dwarfs is crucial for modelling efforts. However, previous investigations have almost exclusively focussed on perturbers as massive as giant or terrestrial planets. Here, through a series of computationally expensive simulations, we have identified the lowest feasible perturbing mass to be about as massive as Earth's moon, $1M_{\leftmoon}$ (Fig. \ref{punch}). Varying properties of the simulated bodies and architectures suggests that this result is robust to within a factor of a few. Consequently, the largely unexplored, but potentially heavily populated, mass region between Luna and the Earth should be considered in future white dwarf pollution studies.

\section*{Acknowledgements}
{\rev We thank the reviewer for helpful comments that have improved the manuscript.} DV gratefully acknowledges the support of the STFC via an Ernest Rutherford Fellowship (grant ST/P003850/1).

\section*{Data Availability}

The simulation inputs and results discussed in this paper are available upon reasonable request to the corresponding author.

\label{lastpage}
\end{document}